\documentclass[12pt]{article}
\usepackage{amsfonts}
\usepackage{latexsym}
\usepackage{amsmath}
\usepackage{amssymb}
\usepackage{amssymb}

\newcommand{\newsection}{    
\setcounter{equation}{0}\section}
\def\appendix#1{\addtocounter{section}{1}\setcounter{equation}{0}
\renewcommand{\thesection}{\Alph{section}}
\section*{Appendix \thesection\protect\indent \parbox[t]{11.15cm}{#1}}
\addcontentsline{toc}{section}{Appendix \thesection\ \ \ #1}}

\newcommand{\be}{\begin{eqnarray}}
\newcommand{\ee}{\end{eqnarray}}
\newcommand{\bea}{\begin{eqnarray}}
\newcommand{\eea}{\end{eqnarray}}
\newcommand{\ba}{\begin{array}}
\newcommand{\ea}{\end{array}}
\newcommand{\nn}{\nonumber \\}

\def \la {\label}

\def\e{\epsilon}

\def\cL{{\cal L}}

\def\hn{{\hat{\nu}}}

\def\bbe{{\bf{e}}}
\font\mybb=msbm10 at 11pt

\def\bb#1{\hbox{\mybb#1}}

\def\bZ {\bb{Z}}
\def\bR {\bb{R}}

\def\ui {{\underline {i}}}
\def\uj {{\underline {j}}}
\def\uk {{\underline {k}}}

\def\hn {{\tilde{\nabla}}}

\begin{document}
\begin{titlepage}
\begin{center}
\vspace{5.0cm}

\vspace{3.0cm} {\Large \bf  Heterotic Black Horizons}
\\
[.2cm]

{}\vspace{2.0cm}
 {\large
J.~Gutowski and  G.~Papadopoulos
 }

{}

\vspace{1.0cm}
Department of Mathematics\\
King's College London\\
Strand\\
London WC2R 2LS, UK\\

\end{center}
{}
\vskip 3.0 cm
\begin{abstract}
We show that the supersymmetric  near horizon geometry of   heterotic black holes is either an
$AdS_3$ fibration over a
 7-dimensional
manifold which admits a $G_2$ structure compatible with a connection with skew-symmetric torsion, or it is a product $\bR^{1,1}\times {\cal S}^8$,
 where ${\cal S}^8$ is a holonomy
$Spin(7)$ manifold, preserving 2 and 1 supersymmetries respectively. Moreover, we demonstrate that the $AdS_3$ class
of   heterotic
horizons can  preserve 4, 6 and 8 supersymmetries provided that the geometry of the base space is further restricted.
Similarly $\bR^{1,1}\times {\cal S}^8$ horizons with extended  supersymmetry are products of $\bR^{1,1}$ with special
holonomy manifolds. We have also found that the heterotic horizons with 8 supersymmetries are locally isometric
to  $AdS_3\times S^3\times T^4$,
 $AdS_3\times S^3\times K_3$ or $\bR^{1,1}\times T^4\times K_3$, where the radii of $AdS_3$ and $S^3$ are equal and the dilaton
 is constant.
\end{abstract}
\end{titlepage}

\setcounter{section}{0}
\setcounter{subsection}{0}


\newsection{Introduction}

It has been known for some time, following  Israel's  uniqueness proof for the Schwarzschild black hole
\cite{israel} and the early results
of  \cite{carter, hawking},
that the most general rotating asymptotically flat black hole solution in four dimensions is
the Kerr solution \cite{robinson1} which is characterized by its mass and angular momentum. These results have also been generalized to
black holes with electric and magnetic charges \cite{israel2, mazur}. Under certain assumptions, four-dimensional black holes exhibit spherical horizon geometry;
for a recent review of these classic results as well as an extended set of references see \cite{robinson}.
In five dimensions, there are no generic black hole uniqueness theorems,
and most of the investigations have focused
either on the supersymmetric case, or on static solutions. It has been shown \cite{reallbh} that the near horizon
geometries of supersymmetric black holes are either the near-horizon geometry of the
BMPV black hole \cite{bmpv}, or $AdS_3\times S^2$,  or $\bR^{1,1}\times T^3$;  and so
the horizon is either a (squashed) $S^3$, or $S^1\times S^2$ or $T^3$, respectively. For the first two cases, the full black hole
solutions, and not just the near horizon geometries,  are known.
In particular, it has been shown in \cite{reallbh} that the only supersymmetric, regular, asymptotically flat
black hole which has the near-horizon BMPV solution as its near horizon geometry is the BMPV black hole
\cite{bmpv}; when the BMPV black hole is static, the near horizon geometry simplifies to $AdS_2 \times S^3$.
It is also known that the supersymmetric black ring found in \cite{ring1} has near-horizon geometry
$AdS_3 \times S^2$ (which is also the near horizon geometry of the black string \cite{gibhtown});
however in this case it is not known if the supersymmetric black ring is the unique solution with this
near-horizon geometry. No black hole solution has been found with
 near horizon geometry\footnote{It is likely that there is no black hole solution with  $\bR^{1,1}\times T^3$ near horizon
 geometry and this background is just the vacuum of a compactification of 5-dimensional supergravity to two dimensions.}
  $\bR^{1,1}\times T^3$. Uniqueness theorems have also been constructed in various theories for static black holes
  in higher dimensions \cite{gibbons1, rogatko}, and for
  black holes  that admit a sufficient number of commuting rotational isometries
  \cite{extr1, extr2, extr3, extr4, extr5}. An effective worldvolume theory for
  higher dimensional black branes has also been recently proposed in \cite{obers1}.

The main aim of this paper is to investigate the near horizon geometry of supersymmetric
heterotic black holes. For this, we shall use the solution of the Killing spinor equations (KSE)
of heteroric supergravity presented in \cite{het1, het2, het3} and adapt the results to the near horizon geometry of
supersymmetric black holes. The latter is described in  a Gaussian null co-ordinate system adapted to
the stationary Killing vector field of a black hole.  The description of this coordinate system as well as the definition
of the near horizon geometry are reviewed in section 2. We shall mostly focus on the near horizon geometries
for which  the 3-form flux $H$ is closed. This is the case whenever the anomalous contribution to the Bianchi identity vanishes. However,
we shall also investigate the geometry of the solutions with $dH\not=0$  and point out the differences between the two cases.

Supersymmetric solutions of heterotic supergravity always admit a null, $\hat\nabla$-parallel, and so Killing,
vector field $V$,
where $\hat\nabla$ is the connection whose skew-symmetric torsion is the 3-form flux of the theory. Moreover, the supersymmetric
solutions for which the holonomy
of $\hat\nabla$ is compact  admit at least one time-like
$\hat\nabla$-parallel, and so Killing, vector field. It is clear from this that all heterotic supersymmetric black holes,
and so their near horizon geometries,
admit at least one null Killing vector field defined everywhere on the black hole spacetime. The compactness of the horizon imposes
additional restrictions on the geometry. In particular, there are near horizon geometries which preserve 2, 4, 6 and
8 supersymmetries. The geometry
of spacetime for this class is a principal bundle $P(G,B;\pi)$ with fibre group $G=SL(2,\bR)\times H$, where
 $H= \{1\}$, $U(1)$ and $SU(2)$,
and the base space $B$ has structure group $G_2$, $SU(3)$  and $SU(2)$, respectively. The structure group of $B$
is compatible with a metric connection $\hat{\tilde \nabla}$ with skew-symmetric torsion. In the cases with
2, 4 and 6 supersymmetries, the $SL(2,\bR)$ subgroup is gauged  over $B$ with a  $U(1)$ connection, and  the fibration
can be geometrically twisted.
The dilaton may not be constant and typically depends on the coordinates of $B$.
However,  in the cases with 8 supersymmetries, the spacetime is a product $AdS_3\times M_7$ and the dilaton is constant.

We furthermore show that all other near-horizon geometries have
a constant dilaton, and the 3-form flux vanishes.
In this case, the near horizon geometries that admit
one supersymmetry
are  isometric to $\bR^{1,1}\times {\cal S}^8$, where ${\cal S}^8$ is a holonomy $Spin(7)$ manifold.
Moreover, there are also solutions $\bR^{1,1}\times {\cal S}^8$ which preserve 2, 3, 4 and 8 supersymmetries
provided
that ${\cal S}^8$ has holonomy $SU(4)$ (Calabi-Yau) and $G_2$, $Sp(2)$ (hyper-K\"ahler),   $\times^2 Sp(1)$ and $SU(3)$
(Calabi-Yau),
and $Sp(1)$ (hyper-K\"ahler)
manifold, respectively.
A more detailed exposition which includes the geometry of the horizons can be found in table 1.

The near horizon geometries with 8 supersymmetries are also classified. We prove that
they are isometric\footnote{If the
dilaton is allowed to be singular on the horizon, $AdS_3\times S^3\times S^3\times S^1$ is also a solution. In particular,
the dilaton
depends linearly on the angular coordinate of $S^1$ and so it is not periodic.}
up to discrete identifications, to
$AdS_3\times S^3\times T^4$, $AdS_3\times S^3\times K_3$ and $\bR^{1,1}\times T^4\times K_3$. In the first two cases,
the fibre group $G=SL(2,\bR)\times SU(2)$, $SL(2,\bR)=AdS_3$ and $SU(2)=S^3$
does not twist over the base space $T^4$ or $K_3$ and the solution is a product.
The radii of $AdS_3$ and $S^3$ are equal, the dilaton is constant and the 3-form field strength is the sum
of the volume forms of the non-abelian groups in the product. Moreover, we demonstrate that $AdS_3\times S^3\times T^4$
does not receive $\alpha'$ corrections.

Our paper is organized as follows. In section 2, we examine the relation between near horizon geometries
and supersymmetry, emphasizing some of the silent features. In section 3, we solve the KSEs
for heterotic horizons which exhibit one superymmetry. In section 4, we show heterotic horizons with non-trivial fluxes
necessarily exhibit at least 2 supersymmetries, and the spacetime admits a $G_2$ structure. In section 5, we analyze the KSEs
 for heterotic horizons with extended supersymmetry, and find that the solutions with non-trivial fluxes
preserve 2, 4, 6 and 8 superymmetries. In sections 6, 7 and 8, we solve the
KSEs for heterotic horizons with non-trivial fluxes and identify the spacetime geometry. In particular in section
8, we classify all heterotic horizons that preserve 8 supersymmetries. In section 9, we first examine the $\alpha'$ corrections
of heterotic horizons.  Then we compare our heterotic horizon geometries
with those that arise in the context of brane configurations, and discuss the regularity of the dilaton. In
section 10, we give our conclusions.
In appendices A and B, we have collected some calculations necessary for the analysis of the KSEs.

\newsection{Near horizon geometry and $N=1$ supersymmetry}

\subsection{ Near horizon limit for extreme black holes}

In what follows, we shall focus on black holes for which the event horizon  is  a Killing horizon, i.e.
there is a  time-like (stationary) Killing vector field $K$  defined  everywhere on the spacetime which becomes null
only on the horizon. It has been shown for non-extremal black holes in higher-dimensional Einstein-Maxwell
theory that the event horizon is a Killing horizon, and furthermore there must exist at least one rotational Killing vector
\cite{symm1}. A similar analysis has been carried out for extremal solutions of Einstein-Maxwell-Dilaton
theory in \cite{symm3}, and modulo certain technical assumptions, the same result holds. However,
for the heterotic theory under consideration here, we shall simply assume that the event horizon is a Killing horizon.
Therefore ${\cal H}$  is identified with the hyper-surface
given by $g(K,K)=0$.
Under this assumption, one can adapt Gaussian null coordinates to  $K$ and the black hole metric near ${\cal H}$
can be written  as \cite{gnull}
\bea
ds^2=2 (dr+r h_I dy^I+r f du) du+\gamma_{IJ} dy^I dy^J~,
\la{nearhor}
\eea
where $K=\partial_u$ and the components
of the metric depend on $r,y$. The $r$ coordinate is chosen such that  ${\cal
H}$ is located at $r=0$. Since $g(K,K)=r f$, $K$ becomes null at the
horizon as expected. Regularity at the horizon requires that at $r=0$ the
metric is non-singular; typically the components of the metric are taken to be analytic in $r$.

The expression (\ref{nearhor}) for the metric
is not unique. If $V$ is any other Killing vector field such that
\bea
g(V,V)\vert_{\cal H}=0~,~~~{\cal L}_V g(K,K)=0~,
\eea
one can introduce a Gaussian null
co-ordinate system adapted to $V$.
The expression
for the spacetime metric is as in (\ref{nearhor}) but now $V=\partial_u$. We shall use
this arbitrariness in adapting a Gaussian null co-ordinate system in the investigation of
supersymmetric black holes later.
The spatial horizon section ${\cal{S}}^8$ given by $u={\rm const}, r=0$,
with metric $ds^2_{\cal{S}} = \gamma_{IJ} dy^I dy^J$, is required to be compact.
The above analysis can also be adapted in the presence of fluxes,
like a Maxwell field, or supergravity form
field strengths.

Since the components of the metric in (\ref{nearhor}) are analytic in $r$,
one has an expansion
\bea
h_I(y,r)&=&\sum^\infty_{n=0} {r^n\over n!} \partial_r^n h_I\vert_{r=0}~,
\cr
f(y,r)&=&\sum^\infty_{n=0} {r^n\over n!} \partial_r^n f\vert_{r=0}~,
\cr
\gamma_{IJ}(y,r)&=&\sum^\infty_{n=0} {r^n\over n!} \partial_r^n\gamma_{IJ}\vert_{r=0}~.
\eea
Some  black hole properties  depend on the first few non-vanishing  terms in the above analytic expansions.
In particular, calculating the surface gravity, one finds that
\bea
i_K dK\vert_{r=0}=-f(y,0) K\vert_{r=0}~,
\eea
where we have used the same symbol to denote the vector field $K$ and the associated 1-form.
Thus if $f(y,0)\not=0$, the black hole has temperature. So for {\it extreme} black holes, one should take $f (y,0)=0$.

To define the {\it near horizon geometry} of an {\it extreme} black hole, we perform the coordinate transformation
\bea
r\rightarrow \epsilon r~,~~~u\rightarrow \epsilon^{-1} u~,~~~~y^I\rightarrow y^I
\la{lim}
\eea
and in the resulting metric, we take the limit $\epsilon\rightarrow 0$. After taking this limit,  the metric reads
 as
\bea
ds^2=2 (dr+ r h_I dy^I+ r^2 \Delta du) du+\gamma_{IJ} dy^I dy^J~,
\la{nearhorb}
\eea
where now $h_I$ and $\gamma_{IJ}$ are evaluated at $r=0$ and $\Delta=\partial_r f\vert_{r=0}$. Observe that if
$f(y,0)\not=0$, the above limit does not exist. Thus the near horizon geometry can only be defined for
extreme black holes.

The near horizon geometry (\ref{nearhorb}) of an extreme black hole admits, in addition to
 $K=\partial_u$,
 an extra
 Killing vector field
 \bea
  D=-r\partial_r+ u \partial_u~,
  \la{dil}
  \eea
associated with the scale symmetry
$
r\rightarrow \ell\, r~,~~~u\rightarrow \ell^{-1}\,  u~
$
which does not commute with $K$, i.e.
\bea
[K,D]= K~.
\eea

In the presence of other fields, like Maxwell or supergravity form field strengths, one can extend the definition of the above limit. In
particular for heterotic supergravity, the theory admits a 2-form gauge potential $B$.  So one has
\bea
B=b du\wedge dr+ b_I dr\wedge dy^I+c_I du\wedge dy^I+b_{IJ} dy^I\wedge dy^J~,
\eea
where all components depend on  $y^I$ and $r$ coordinates.
Assuming analyticity in the components of $B$ in the $r$ coordinate, one can define the near horizon gauge potential
by taking the limit (\ref{lim}) provided that $c_I(y,0)=0$. This condition is similar to the extremality restriction
for the metric. After taking the limit, the gauge potential can be rearranged as
\be
B = r\, du \wedge N + S\, du \wedge (dr+ r h_Idy^I) + W~,
\ee
where now $S$ is a scalar function, and  $N$ and $W$ are 1- and 2-forms  on ${\cal S}^8$, respectively.
The $b_I$ component of $B$ vanishes in the limit.
Observe that $B$ is also invariant under the scale symmetry generated by Killing vector (\ref{dil}). One therefore
concludes that the scale symmetry is a {\it generic} feature of the near horizon geometries. The dilaton $\Phi$ in the near horizon limit
depends only on the $y$ coordinates.

{}For later use, we collect the heterotic fields in the near horizon limit  as
\bea
ds^2&=&2 \bbe^+ \bbe^- + \delta_{ij} \bbe^i \bbe^j~,
\cr
H &=&  \bbe^+ \wedge \bbe^- \wedge \big(dS-N-Sh\big)
+ r \bbe^+ \wedge \big(h \wedge N - dN - S dh \big) + dW~,
\cr
\Phi&=&\Phi(y)~,
\la{bhdata}
\eea
where $H:=dB$,
\bea
\bbe^+ = du~,~~~
\bbe^- = dr + r h+r^2\Delta du ~,~~~
\bbe^i &=& e^i{}_I dy^j~,
\label{nhbasis}
\eea
and $\gamma_{IJ} =\delta_{ij} \, e^i{}_I e^j{}_J$.

\subsection{Supersymmetry}

The supersymmetric heterotic backgrounds have been classified in \cite{het1, het2}.  There are two large classes of
supersymmetric heterotic backgrounds depending on whether  the holonomy
of the connection, $\hat\nabla$,
with skew-symmetric torsion $H$,  is a subgroup of a compact or non-compact isotropy group of $\hat\nabla$-parallel spinors. We shall first focus on the non-compact case and in particular
on the backgrounds which preserve one supersymmetry. For these backgrounds the holonomy of $\hat\nabla$ is contained
in $Spin(7)\ltimes\bR^8$ and  admit a local frame $(e^+, e^-, e^i)$
such that
\bea
\hat\nabla e^-=0~,~~~\hat\nabla(e^-\wedge \phi)=0~,~~~
\eea
where
\bea
\phi={1\over4!} \phi_{i_1\dots i_4} e^{i_1}\wedge \dots\wedge e^{i_4}~,
\eea
is the self-dual fundamental form of $Spin(7)$. This is the full content of the gravitino KSE. The dilatino
KSE implies that
\bea
\partial_+\Phi=0~,~~~de^-\in \mathfrak{spin}(7)\oplus \bR^8~,~~~2\partial_i\Phi=(\theta_\phi)_i+H_{-+i}~,
\la{dcon}
\eea
where $\theta_\phi =-{1\over6}\star(\star \tilde d\phi\wedge \phi)$ is the Lee form of $\phi$ and
$\tilde d$ is the exterior derivative projected along the directions transverse to the light-cone. For a detailed  explanation of these results and for
our notation, see \cite{het1, het2}.

The metric and 3-form field strength can then be expressed as
\bea
ds^2&=&2 e^- e^++\delta_{ij}\, e^i e^j
\cr
H&=&d(e^-\wedge e^+)+ e^-\wedge L+\tilde H~,~~~\tilde H={1\over3!} H_{i_1i_2i_3} e^{i_1}\wedge e^{i_2}\wedge e^{i_3}~,
\la{susydata}
\eea
where $L\in\mathfrak{spin}(7)$ is not determined by the KSEs and $\tilde H$ can be expressed
in terms of $\phi$ and its first derivatives as
\bea
\tilde H=-\star \tilde d \phi+\star(\theta_\phi\wedge \phi)~.
\la{hsusy}
\eea
The expression for $\tilde H$ is as that for 8-dimensional manifolds with $Spin(7)$-structure and compatible connection
with skew-torsion \cite{ivanovx}. This is the full content of the KSEs.

\subsection{Supersymmetric heterotic black holes}

Supersymmetric black holes are those black holes  of supergravity theories that also satisfy the KSE,
and the Killing spinor vector bilinear(s) are well-defined everywhere on the spacetime, and in particular {\it analytic}\footnote{This is an assumption
and it may not follow
from the KSEs of supergravity theories.} in $r$ near the horizon.

Suppose that $V$ is a Killing vector field constructed as spinor bi-linear.
If $g(V,V)=0$ at ${\cal{H}}$ and the black hole is {\it extreme}, using $g(K,K)=0$,
one can show that
\bea
{\cal L}_V g(K,K)\vert_{\cal H}=0~,
\la{vkk}
\eea
i.e. a Gaussian null coordinate system can be adapted to $V$ as well and
the metric can be written as in  (\ref{nearhor}).

Now let us turn to the heterotic case. Supersymmetric backgrounds in heterotic theory admit always a null $\hat\nabla$-parallel,
and so Killing, vector field constructed as a Killing spinor bi-linear. Depending on the number of supersymmetries and the holonomy
of $\hat\nabla$, they may admit a time-like and several space-like $\hat\nabla$-parallel,
and so Killing, Killing spinor vector bi-linears. The time-like and spacelike $\hat\nabla$-parallel vector fields cannot be used to adapt
a null Gaussian coordinate system for a black hole. This is because their length is constant and so they do not vanish anywhere
on the spacetime. So it remains to consider the null vector bilinears $V$. Since $V$ is null, $g(V,V)=0$ everywhere on the spacetime
and so on the horizon as well. Moreover for extreme black holes one also has (\ref{vkk}) and so a Gaussian null
coordinate system can be adapted to $V$. For such a system $V=\partial_u$ is null and so one has $f=0$. If in addition, we take the
near horizon limit now adapted to $V$, the heterotic fields are given in
(\ref{susydata}) and (\ref{hsusy}) but now with $\Delta=0$.

It is not apparent what kind of supersymmetric black holes one should expect to be present in heterotic supergravity.
Asymptotically flat black holes with fluxes flowing over an 8-sphere at infinity
 may be ruled out because of the
presence of an everywhere null Killing vector field which suppresses the dependence on a radial direction that it is needed for the
appropriate decay of the fields.

The class of superymmetric black holes that is expected to be present consists of
 the Kaluza-Klein (KK) black holes. It may seem worrying that if KK black holes are
solutions of heterotic supergravity, there is not a time-like
Killing vector field constructed from Killing spinor bilinears which becomes null at the horizon. However, this is not necessary.
Some cases are known for which, after lifting
a black hole solution from a lower-dimensional theory to 10 or 11 dimensions, the stationary time-like Killing vector field
becomes null everywhere on the spacetime. This happens, for example, when we  lift  the  asymptotically
$AdS_5$ black hole of 5-dimensional supergravity \cite{realgutbh}  to IIB
supergravity. Note though that our assumptions in 10 dimensions require that the spacetime admits a time-like Killing vector field $K$
which becomes null at the horizon. However $K$ may not be written as a Killing spinor bilinear. Nevertheless it is an additional restriction on the geometry of such black hole spacetimes.

We shall solve the heterotic KSEs at the near horizon limit by adapting a
Gaussian null coordinate system
to the null Killing vector field constructed from a Killing spinor bilinear. We shall refer to all these solutions as
{\it heterotic black horizons} or simply horizons. However, it is not apparent that all these
geometries can be extended to a black hole spacetime. It is
likely that some of them are simply the Kaluza-Klein vacua of compactifications of
 heterotic supergravity to two dimensions.
Nevertheless the heterotic horizons include all the near horizon geometries of extreme supersymmetric heterotic black holes.

\newsection{N=1 heterotic  horizons}

\subsection{N=1 supersymmetry}

As we have explained in the previous section, we adapt Gaussian  null coordinates to the vector field
constructed as a  bilinear of the Killing spinor of these backgrounds. In general, the natural frame adapted
to supersymmetric backgrounds $(e^+, e^-, e^i)$ is distinct to that associated with the Gaussian null
coordinates $(\bbe^+, \bbe^-, \bbe^i)$. But, as we have explained,  the  Gaussian null coordinates are taken with respect
to the null Killing spinor bilinear, $\bbe^-=e^-$. Moreover, we shall show in appendix A that without loss
of generality, one can set
\bea
e^-=\bbe^-~,~~~e^+=\bbe^+~,~~~e^i=\bbe^i~;~~~\Delta=0~,
\la{ebbe}
\eea
where $\Delta$ vanishes because the Killing vector field is null.
Comparing the expression for $H$ in (\ref{susydata}) and (\ref{bhdata}),
one finds that
\bea
ds^2&=& 2 \bbe^- \bbe^++ d\tilde s^2~,~~~ d\tilde s^2=\delta_{ij} \bbe^i \bbe^j~,~~~
\cr
H&=&d(\bbe^-\wedge \bbe^+)+\tilde H~,~~~\tilde H=dW~,
\cr
\Phi&=&\Phi(y)~,
\la{shh}
\eea
where
\bea
\bbe^-=dr+r h_i e^i~,~~~ \bbe^+=du~,~~~\bbe^i=e^i{}_Idy^I~,
\eea
see also (\ref{nhbasis}), and the Killing spinor is
\bea
\e=1+e_{1234}~.
\eea
It is essential in what follows to notice that $h$ is a {\it globally} defined 1-form on the horizon section ${\cal S}^8$.
Observe that the term involving $L$ in (\ref{susydata}) vanishes in the near horizon limit.

The conditions that arise in the dilatino KSE (\ref{dcon}) can be rewritten as
\bea
dh\in \mathfrak{spin}(7)~,~~~2\partial_I\Phi+h_I=(\theta_\phi)_I~.
\label{heq}
\eea
The only condition that the gravitino KSE imposes is that the submanifold ${\cal S}^8$ given by $r=u={\rm const}$ has a  $Spin(7)$
structure.
This is because every  manifold with a $Spin(7)$ structure admits a compatible connection with skew-symmetric torsion
given in (\ref{hsusy}). In particular, one has that
\bea
\hat{\tilde\nabla}_I\phi_{J_1\dots J_4}=0
\label{spneq}
\eea
where $\hat{\tilde\nabla}$ is the connection with skew-symmetric torsion constructed from the data $(d\tilde s^2, \tilde H)$ of ${\cal S}^8$, and
$\phi$ is the fundamental self-dual 4-form of $Spin(7)$.
However,  the closure of $\tilde H$ given in (\ref{hsusy}) is not implied by the KSEs and has to be imposed as an additional constraint.

\subsection{Field equations}

It is known that the KSEs, the  field equations of the 3-form flux,  the  $E_{--}$
component of the Einstein equations  and $dH=0$ imply all the rest of the field equations of heterotic supergravity \cite{het1, het2}.
So to find the heterotic horizons, we
have to solve in addition to the KSEs some of the field equations of the theory.
Using the special geometry of heterotic horizons (\ref{shh}),  we decompose the field equation of the 3-form flux
\be
d \star \big( e^{-2 \Phi} H  \big) =0
\ee
in terms of the various forms defined on ${\cal{S}}^8$ as
\be
\label{geq1}
\tilde \nabla_i (h^i e^{-2\Phi})=0~,~~~
\ee
\be
\label{geq2}
e^{2\Phi} \tilde \nabla_j \big (e^{-2\Phi} (dh)^{ji}\big)+{1\over2} (dh)_{jk} \tilde H^{jki}+ h_j (dh)^{ji}=0~,
\ee
\be
\label{geq3}
e^{2\Phi} \tilde \nabla_k \big( e^{-2\Phi} \tilde H^{kij}\big)+ (dh)^{ij}- h_k \tilde H^{kij}=0~,
\ee
where here the frame indices are those of ${\cal{S}}^8$, and $\tilde \nabla$ is the Levi-Civita connection of $d\tilde s^2$. In what follows, we shall also use the field equation of
the dilaton
\be
\label{deq}
\tilde \nabla^2 \Phi - h^i \tilde \nabla_i \Phi -2 \tilde \nabla^i \Phi \tilde \nabla_i \Phi +{1 \over 12} \tilde H_{ijk} \tilde H^{ijk} -{1 \over 2} h_i h^i =0~,
\la{dfe}
\ee
and the $E_{ij}$ component of the Einstein equations
\be
{\tilde{R}}_{ij} = {1 \over 4} \tilde H_{imn} \tilde H_j{}^{mn} -2 \tilde \nabla_i \tilde \nabla_j \Phi
- \tilde \nabla_{(i} h_{j)}~,
\la{efe}
\ee
where ${\tilde{R}}_{ij}$ denotes the Ricci tensor of ${\cal S}^8$.

\subsection{Solutions}

\subsubsection{$h=0$}

We first consider solutions for which $h=0$. In this case, the dilaton field equation (\ref{dfe})  can be written as
\be
\tilde\nabla^2 e^{-2 \Phi} = {1 \over 6} e^{-2 \Phi} \tilde H_{ijk} \tilde H^{ijk} \ .
\ee
Hence, the maximum principle implies that $\Phi$ is constant and $\tilde H=0$.
It follows that $H=0$ and the spacetime metric
is
\be
ds^2 = ds^2 (\bR^{1,1}) + ds^2({\cal S}^8)~.
\ee
Moreover  ${\cal S}^8$ is a compact holonomy  $Spin(7)$ manifold \cite{berg}. Examples of such manifolds can be found   in \cite{joyce}.
Such geometries are also the vacua of heterotic compactifications to two dimensions.

Before we proceed to examine the remaining cases, it is worth mentioning that the dilaton field equation together with compactness
impose strong restrictions on the existence of solutions. This is the case irrespective of whether the solution is supersymmetric or not,
but it is dependent on the couplings. For this consider solutions with metric
\bea
ds^2=ds^2(\bR^{n,1})+ds^2(X)~,
\eea
  for which $H$ is either purely magnetic or purely electric, and all fields  depend only on the coordinates of $X$.
  If $X$ is compact, then it is clear that the dilaton field equation implies that only solutions  are those
  that have  constant dilaton, $H=0$ and $X$ Ricci-flat. This is in agreement with the
  more general results in \cite{gibbons}. A similar conclusion can be reached using the KSEs
  \cite{stefan}.

\subsubsection{ $\tilde H=0$}

Suppose that $\tilde H=0$. In this case,  (\ref{spneq}) implies that ${\cal S}^8$ is a holonomy  $Spin(7)$ manifold. As a result
the Lee form $\theta$ vanishes and the dilatino KSE implies that
\be
\label{closh}
h=-2 d \Phi \ .
\ee
Substituting this condition into the dilaton equation ({\ref{deq}}), one obtains
\be
\tilde\nabla^2 \big(e^{-2 \Phi}\big) =0~.
\ee
Again compactness of ${\cal S}^8$ and the maximum principle implies that $\Phi$ is constant.
In turn, ({\ref{closh}}) gives  $h=0$. Thus the solutions we find
are identical to those of the previous section.

\subsubsection{ $h \neq 0$ and $\tilde H \neq 0$}

It remains to investigate the heterotic horizons  for which both $h$ and $\tilde H$ are non-vanishing. It turns out
that such solutions always preserve at least two supersymmetries,  and both the spacetime $M$ and ${\cal S}^8$ admit a $G_2$ structure. We examine this case in the following section.

\newsection{ The $G_2$ structure of heterotic horizons}

Let us assume that $h$ and $\tilde H$ do not vanish and that the heterotic horizons admit one supersymmetry.
To prove that such heterotic horizons admit two supersymmetries and that the holonomy of the connection $\hat\nabla$ reduces to $G_2$,
 we shall  show  first that the $Spin(7)$ holonomy
of $\hat{\tilde \nabla}$ reduces to $G_2$.

\subsection{$\hat{\tilde \nabla}$ has $G_2$ holonomy}

To proceed with the analysis, we shall first
compute the Laplacian of $h^2$, where $h^2= h_i h^i$. In particular, we find the identity

%
%
%

\be
\label{comp1}
\hn^2 h^2 +(-2 d \Phi + h)^j \hn_j h^2 = 2 \hn_{(i} h_{j)} \hn^{(i} h^{j)}
+{1 \over 2} (dh-i_h \tilde H)_{ij} (dh -i_h \tilde H)^{ij}~.
\ee
To obtain this expression, write
\bea
\hn^2 h^2=2 \hn_i h_j \hn^i h^j+ 2 h^j \hn^2 h_j=2 \hn_i h_j \hn^i h^j+2\hn^i (dh)_{ij} h^j+2 \tilde R_{ij} h^i h^j+2 h^j \tilde \nabla_j (\tilde \nabla_i h^i)~.
\nonumber \\
\eea
We then use the field equation ({\ref{geq1}}) to eliminate the
$\tilde \nabla_i h^i$ term in favour of  $2h^i \tilde \nabla_i \Phi$,
and then use (\ref{geq2}) and (\ref{efe}).
After some re-arrangement of terms,
one obtains (\ref{comp1}).

Applying the maximum principle on ({\ref{comp1}}) using the compactness of ${\cal S}^8$, we find that
$h^2$ must be constant, and hence the RHS of ({\ref{comp1}}) must vanish identically.
Therefore, we have that
\be
\label{new1}
\hn_{(i} h_{j)}=0~,~~~dh = i_h \tilde H~.
\ee
These two conditions are equivalent to requiring that
\bea
\hat{\tilde \nabla} h=0~.
\la{new2}
\eea
Thus $h$ is a $\hat{\tilde \nabla}$-parallel vector of ${\cal S}^8$.
Since the isotropy group of a non-vanishing  element in the spinor representation of   $Spin(7)$ holonomy is $G_2$, the holonomy of $\hat{\tilde \nabla}$
 is contained in  $G_2$.

Using the above results, one can  show the identities
\be
\label{liex1}
i_h dh =0~,
\ee
\be
\label{liex2}
\cL_h \tilde H=0~,
\ee
\be
\label{liex2b}
\cL_h \Phi =0~,
\ee
and
\be
\label{liex3}
\cL_h \phi =0~.
\ee
The first identity follows from the properties  that $h^2$ is  constant and $h$ is Killing,  the second  follows from the second condition
in (\ref{new1}) and $d\tilde H=0$, the third follows from
({\ref{geq1}}), and the fourth follows from the first equation in (\ref{heq}) and (\ref{spneq}).

\subsection{Killing spinor equations revisited}

Before we proceed to prove that the spacetime admits an additional supersymmetry, it is convenient to reexamine
the KSEs of the backgrounds assuming that they admit at least one supersymmetry, using
the results of the previous section. Suppose that $\e$ is a Killing spinor. The gravitino KSE implies that $\e$ does not depend on $r$. Moreover
\bea
\e= {u\over2} h_i \Gamma^{-i} \eta_-+\eta_++\eta_-~,~~~\Gamma_\pm \eta_\pm=0~,
\la{klis}
\eea
where $\eta_\pm=\eta_\pm(y)$, solves the gravitino KSE, iff
\bea
\hat{\tilde \nabla}\eta_\pm=0~,
\label{ax3}
\eea
\bea
dh_{ij}\Gamma^{ij}\eta_\pm=0~.
\label{ax1}
\eea
In addition $\e$ solves the dilatino KSE, iff
\be
\label{ax2}
\bigg( \big(2 d \Phi \mp h \big)_i \Gamma^i -{1 \over 6} \tilde H_{ijk} \Gamma^{ijk} \bigg)  \eta_\mp=0~.
\ee
Furthermore, it suffices to solve (\ref{ax3})-(\ref{ax2})  for either  $\eta_-$ or $\eta_+$. Notice that if there is a solution
$\eta_+$, then there is another solution with $\eta_-=\Gamma^{+i}h_i \eta_+$, and vice versa. This is because $\Gamma^{+i} h_i$
and $\Gamma^{-i} h_i$ commute with (\ref{ax3}) and (\ref{ax1}), and anti-commute with  (\ref{ax2}) up to a change of sign in the $h$ term.
One can demonstrate this
by using the relations (\ref{new1}), (\ref{new2}) and (\ref{liex1}) of the previous section.
This will simplify the analysis for all heterotic horizons that admit more than one supersymmetry.

\subsection{N=2 supersymmetry and  $G_2$ holonomy}

To construct the second Killing spinor, we set $\eta_+=1+e_{1234}$. This spinor satisfies the KSEs (\ref{ax3})-(\ref{ax2}) because
these are the conditions that arise on the geometry from the requirement that the solutions admit one supersymmetry.
Moreover, we set $\eta_-=h_i\Gamma^{+i} \eta_+$ and substitute this into (\ref{klis}) to find that the two linearly independent Killing spinors are
\bea
\epsilon^1=1+e_{1234}~,~~~
\epsilon^2=- k^2 u (1+e_{1234})+h_i \Gamma^{+i}  (1+e_{1234})~,
\la{g2ks}
\eea
where $k^2=h^2$ is the constant length of $h$.
It is easy to see from the results of \cite{het1} that the isotropy group of both Killing spinors in $Spin(9,1)$ is $G_2$. Therefore
the holonomy of $\hat\nabla$ reduces to a subgroup of $G_2$.

\subsection{Geometry}

\subsubsection{Geometry of spacetime}

To investigate the geometry of spacetime, we first compute the 1-form bi-linears $\lambda$ associated with the Killing spinors
(\ref{g2ks}) to find
\bea
\lambda^- &=& \bbe^-~,~~~
\lambda^+ = \bbe^+ - {1 \over 2} k^2 u^2 \bbe^- -u h~,~~~
\lambda^1 = k^{-1} \big(h+ k^2 u \bbe^-\big)~.
\la{g2vbi}
\eea
Moreover, the associated vector fields $\xi_a$, $a=-,+1$, satisfy the Lie bracket algebra
\be
[\xi_+, \xi_-] = -k \xi_1, \qquad  [\xi_+, \xi_1]=k \xi_+, \qquad  [ \xi_- , \xi_1]= -k \xi_-~,
\la{2lie}
\ee
which is isomorphic to $\mathfrak{sl}(2,\bR)$. It is then a consequence of the classification results of \cite{het1} that the spacetime
is a principal bundle $M=P(SL(2, \bR), B^7; \pi)$ with fibre group $SL(2,\bR)$,  base space $B^7$ and principal bundle connection
$\lambda$. Moreover the spacetime metric
and 3-form flux can be written as
\bea
ds^2&=&\eta_{ab} \lambda^a \lambda^b+d\tilde s_{(7)}^2~,~~~
H=CS(\lambda)+\tilde H_{(7)}~,
\eea
where $d\tilde s_{(7)}^2$ and $\tilde H_{(7)}$ is a metric and 3-form flux of $B^7$, respectively, ie
$d\tilde s_{(7)}^2$ and $\tilde H_{(7)}$ are orthogonal to the directions $\lambda^a$. Moreover
\bea
CS(\lambda)={1\over3} \eta_{ab} \lambda^a\wedge  d \lambda^b+{2\over3} \eta_{ab} \lambda^a \wedge {\cal F}^b~,
\eea
is the Chern-Simons form of $\lambda$, where
\be
{\cal{F}}^a = d \lambda^a - {1\over2}H^a{}_{b_1 b_2} \lambda^{b_1} \wedge \lambda^{b_2}~,
\ee
is the curvature of $\lambda$. The dilaton $\Phi$ depends only on the coordinates of $B^7$.

The curvature of $\lambda$ is non-vanishing and so the  $SL(2,\bR)$ fibre twists over the base space $B^7$. In particular,
one finds that
\bea
\label{curvnh1}
{\cal{F}}^+ &=& -u(1+{1 \over 2}k^2 r u) dh~,
\nn
{\cal{F}}^- &=& r dh~,
\nn
{\cal{F}}^1 &=& k^{-1}(1+k^2 r u) dh~.
\la{sl2rc}
\eea
Moreover a straightforward calculation reveals that
\bea
CS(\lambda)=du\wedge dr\wedge h+r du\wedge dh+k^{-2} h\wedge dh~.
\eea

Since ${\cal F}$ has only one independent component determined by $dh$, it is clear that only an abelian
subgroup of the $SL(2,\bR)$ fibre is gauged. As expected ${\cal F}^a$ is a 2-form over $B^7$, because $i_h dh=0$,
and a $G_2$-instanton, i.e.
\bea
dh\in \mathfrak{g}_2~.
\la{g2inst}
\eea
This can  been seen from (\ref{ax1}). The $G_2$ fundamental form is
\bea
\varphi=k^{-1} i_h \phi~,
\eea
where $\phi$ is the fundamental $Spin(7)$ form.
The  conditions that arise from the dilatino KSE (\ref{ax2}) are
\bea
k-{1\over6} H_{ijk}\varphi^{ijk}=0~,~~~\theta_\varphi=2d\Phi~,
\la{g2d}
\eea
where now $i,j,k=2,3,4,6,7,8,9$.

The geometric data $(d\tilde s_{(7)}^2, \tilde H_{(7)})$ of $B^7$ are compatible with a $G_2$ structure, i.e.
the metric connection, $\hat{\tilde \nabla}^{(7)}$,  on $B^7$ with skew-symmetric torsion $\tilde H_{(7)}$
has holonomy contained in $G_2$. This in turn determines $\tilde H_{(7)}$ in terms of the fundamental $G_2$ form
$\varphi$
as
\bea
\tilde H_{(7)}=k  \varphi+ e^{2\Phi} \star_7d\big( e^{-2\Phi} \varphi\big)~.
\la{g2H}
\eea
In contrast to the $Spin(7)$ case, not all 7-dimensional manifolds with a $G_2$ structure
 admit a compatible connection with skew-symmetric torsion \cite{ivanovy}. For this to hold, one must in addition have
$d[e^{-2\Phi}\star_7\varphi]=0$.

Furthermore, in the $G_2$ holonomy case all the field equations of the heterotic supergravity are implied provided that $dH=0$.
Therefore,  the only remaining equations that one has to solve  are
\bea
d[e^{-2\Phi}\star_7\varphi]=0~,~~~k^{-2}\,dh\wedge dh+ d\tilde H_{(7)}=0~,~~~
(dh)_{ij}={1\over2} \star_7\varphi_{ij}{}^{kl}
(dh)_{kl}~.
\la{totg2}
\eea
The first is the geometric condition\footnote{The geometric condition can also be written as $d\star_7\varphi=\theta_\varphi\wedge \varphi$
and this corrects a sign in \cite{het1, het2}.} on the $G_2$ structure of $B^7$,  the second arises from $dH=0$ and the
last is equivalent to $dh\in \mathfrak{g}_2$  (\ref{g2inst}). The geometric condition
implies that the manifold $B^7$ must be conformally co-calibrated. The $dH=0$ condition
is more involved  and it is reminiscent of
the equations that one solves for heterotic supergravity after taking into account the
one-loop anomalous contribution. Of course, we have assumed that the anomaly cancels since we have taken $dH=0$.  But from the
perspective of the base space $B$, the equation that $\tilde H$ obeys is similar to that which would hold if there were an anomalous contribution.  We have not been
able to prove that it admits non-trivial solutions. To summarize, the spacetime metric and 3-form field strength can be written as
\bea
ds^2&=&\eta_{ab} \lambda^a \lambda^b+d\tilde s_{(7)}^2~,~~~
\cr
H&=&du\wedge dr\wedge h+r du\wedge dh+k^{-2} h\wedge dh+k  \varphi+ e^{2\Phi} \star_7d\big( e^{-2\Phi} \varphi\big)~,
\eea
subject to the conditions (\ref{totg2}).

\subsubsection{Geometry of ${\cal S}^8$}

The geometry of ${\cal S}^8$ can be investigated separately from that of the spacetime. This is because
the geometry of the KSEs (\ref{ax3})-(\ref{ax2}) can be analyzed without reference to
the original 10-dimensional spacetime. To proceed from now on we shall reserve the Latin indices $i,j,k$ for the base space $B$ in each case
and denote the indices of directions transverse to the light-cone, and so also those of the horizon section ${\cal S}^8$, with $\ui,\uj,\uk$. In this notation
 $\hat{\tilde \nabla}$-parallel spinors
are
\bea
\eta_+^1=1+e_{1234}~,~~~\eta_-^2=\Gamma^{+\ui}h_\ui (1+e_{1234})~.
\la{g2sks}
\eea
The isotropy group of both spinors in $Spin(8)$ is $G_2$ and so the holonomy of $\hat{\tilde \nabla}$ is contained in $G_2$.
The associated $\hat{\tilde \nabla}$-parallel bilinears are
\bea
h~,~~~\varphi~,
\eea
where $\varphi=k^{-1} i_h\phi$ and $\phi$ is the fundamental $Spin(7)$ form. As in the spacetime case, $h$ can be viewed as
the connection of a $S^1$ bundle over a base space $B^7$. The conditions that arise from the dilatino KSE
are given in (\ref{g2d}). The  metric and torsion  of ${\cal S}^8$
can be written as
\bea
d\tilde s^2=k^{-2} h\otimes h+d\tilde s_{(7)}^2~,~~~\tilde H= k^{-2}  h\wedge dh+\tilde H_{(7)}~,
\eea
where $d\tilde s_{(7)}^2$  and $\tilde H_{(7)}$ are given in previous section, see eg (\ref{g2H}). Moreover, they satisfy
(\ref{totg2}).

\newsection{Extended supersymmetry}

\subsection{h=0}

There are two cases to consider depending on whether $h$ vanishes. If $h=0$, we have seen that the spacetime
is a product $\bR^{1,1}\times {\cal S}^8$, where ${\cal S}^8$ is a holonomy $Spin(7)$ manifold. The flux $H$ vanishes and the dilaton is
constant. Substituting these into the KSEs, they reduce to a parallel transport equation for the
Levi-Civita connection $\nabla$ of ${\cal S}^8$. As a result, the solutions with more than one supersymmetry are products, up to discrete identifications, of Minkowski space $\bR^{1,1}$ with
those Berger type of manifolds which admit parallel spinors. The results are tabulated in table 1.

\begin{table}
\centering
\fontencoding{OML}\fontfamily{cmm}\fontseries{m}\fontshape{it}\selectfont
\begin{tabular}{|c|c|c|c|}\hline
$$M$$& ${\rm hol}(\nabla)$ &$N$&${\rm Spinors}$
 \\
\hline\hline
 $\bR^{1,1}\times {\cal S}^8$ & $Spin(7)$ & $1$ & $1+e_{1234}$
 \\
 \hline
$\bR^{1,1}\times {\rm CY}^8$&$SU(4)$&$2$ &$1$
\\
\hline
$\bR^{1,1}\times {\rm  HK}^8$&$Sp(2)$&$3$&$1,  i(e_{12}+e_{34})$
\\
\hline
$\bR^{1,1}\times K_3\times K_3$&$\times^2 Sp(1)$&$4$&$1,  e_{12}$
\\
\hline
$\bR^{1,1}\times S^1\times {\cal S}^7$&$G_2$&$2$&$1+e_{1234}, e_{15}+e_{2345}$
\\
\hline
$\bR^{1,1}\times T^2\times {\rm  CY}^6$&$SU(3)$&$4$&$1, e_{15}$
\\
\hline
$\bR^{1,1}\times T^4\times K_3$&$Sp(1)$&$8$&$1, e_{15}, e_{12}~, e_{25}$
\\
\hline
$\bR^{1,1}\times T^8$&$\{1\}$&$16$&${\rm all}$
\\
\hline
\end{tabular}
\label{tab2}
\begin{caption}
{\small {\rm ~~Some geometric data of the horizon geometries with $h=0$ are described. In the first column, we give  the different spacetime geometries that occur. In the second column, we present  the holonomy groups
of the associated spacetime Levi-Civita connection. In the third and fourth  column, we  describe the number of parallel spinors and  representatives
of the parallel spinors, respectively.  HK and CY stand for hyper-K\"ahler and  Calabi-Yau manifolds, respectively. $T^n$ is the n-dimensional torus.
}}
\end{caption}
\end{table}

\subsection{h$\not=$0}

If $h\not=0$, it suffices to investigate the KSEs (\ref{ax3})-(\ref{ax2}) for the $\eta_+$ spinors. This is because
as we have already mentioned the solutions for the $\eta_-$ spinors are given by $\eta_-=h_\ui\Gamma^{+\ui}\eta_+$. As a result, the heterotic horizons with
$h\not=0$ always preserve an even number of supersymmetries.

\subsubsection{Gravitino}

To solve the gravitino KSE for $\eta_+$, i.e. the parallel transport equation for $\hat{\tilde \nabla}$ (\ref{ax3}), it suffices
to find the subgroups of $Spin(8)$ which leave invariant spinors in  the even chirality Majorana representation of $Spin(8)$.  These are
\bea
Spin(7)~(1), ~~SU(4)~(2), ~~Sp(2)~ (3),~~\times^2 Sp(1)~ (4),  ~~ Sp(1)~(5)~,~~ U(1)~(6),~~\{1\}~(8) \cr
\la{isonew}
\eea
and have been stated in \cite{berg, bryant1, jose1}, where
the number $N^+$ of invariant $\eta_+$ spinors is in parentheses. The number of parallel  spinors of spacetime
is $N=2 N^+$. In particular in the $Sp(1)$ and $U(1)$ cases, the number of Killing spinors is 10 and 12,
respectively. All backgrounds with 10 and 12 supersymmetries are plane waves which in addition are group manifolds \cite{jose2, het2}. Such solutions
do not have $AdS_3$ as a submanifold  and so they must be excluded.

Since both $\eta_+$ and $\eta_-=h_\ui\Gamma^{+\ui}\eta_+$ solve the parallel transport equation for $\hat{\tilde \nabla}$ (\ref{ax3}), the
holonomy of $\hat{\tilde \nabla}$ reduces to the isotropy group of both $\eta_+$ and $\eta_-$ spinors. In particular, it reduces
to the subgroups of (\ref{isonew}) which in addition preserve the parallel vector $h$. It is straightforward to find all these
groups and the results are tabulated in table 2.

\begin{table}
\centering
\fontencoding{OML}\fontfamily{cmm}\fontseries{m}\fontshape{it}\selectfont
\begin{tabular}{|c|c|c|c|}\hline
${\rm Iso}(\eta_+)$& ${\rm hol}(\hat\nabla)$ &$N$&$ \eta_+$
 \\
\hline\hline
  $Spin(7)\ltimes\bR^8$ & $G_2$ & $2$ & $1+e_{1234}$
 \\
 \hline
$SU(4)\ltimes\bR^8$&$SU(3)$&$4$ &$1$
\\
\hline
$Sp(2)\ltimes\bR^8$&$SU(2)$&$6$&$1,  i(e_{12}+e_{34})$
\\
\hline
$\times^2Sp(1)\ltimes\bR^8$&$SU(2)$&$8$&$1,  e_{12}$
\\
\hline
\end{tabular}
\label{tab1}
\begin{caption}
{\small {\rm ~~Some of the geometric data used to solving the gravitino KSE are described.
In the first column, we give the isotropy groups, ${\rm Iso}(\eta_+)$, of $\{\eta_+\}$ spinors in $Spin(9,1)$. In the second column
we state the holonomy of the connection with torsion $\hat\nabla$. The holonomy of $\hat{\tilde\nabla}$ is identical to that of $\hat\nabla$.
In the third column, we present the number of $\hat\nabla$-parallel spinors
and in the last column we give representatives of the $\{\eta_+\}$ spinors.
}}
\end{caption}
\end{table}

\subsubsection{Dilatino}

As in the general heterotic case, only some of solutions of the gravitino KSE (\ref{ax3}) are also solutions of the dilatino one (\ref{ax2}). To find the solutions
of the dilatino KSE given a solution of the gravitino KSE, it suffices to focus on the $\eta_+$ parallel spinors. Since the isotropy groups
of $\eta_+$ spinors in $Spin(9,1)$, given in table 2,  are non-compact, the analysis of the dilatino KSE (\ref{ax2}) is similar to that in \cite{het2} for backgrounds
for which ${\rm hol}(\hat\nabla)$ is a non-compact group. In particular, one can find the $\Sigma$-groups in each case and investigate the orbits
of these on the space of $\{\eta_+\}$ spinors. The end result is that it suffices to investigate only those cases for which
{\it all} $\{\eta_+\}$ $\hat{\tilde\nabla}$-parallel spinors also solve the dilatino KSE. This is because all the rest are just special cases. For example,
in the $SU(4)\ltimes \bR^8$ case in table2, there are two $\hat{\tilde\nabla}$-parallel spinors. It is possible that
only a linear combination of them solves the dilatino KSE, i.e. there is one Killing spinor. If this is the case,
 then  such backgrounds will be included
in those of  $Spin(7)\ltimes\bR^8$,  table 2, for which   the  $\hat{\tilde\nabla}$-parallel spinor also solves the dilatino KSE.
A similar conclusion holds for all the other cases.

\newsection{N=4 horizons}

\subsection{Geometry of spacetime}

The first two Killing spinors are those of the $G_2$ case (\ref{g2ks}). The additional two  Killing spinors  are given by
\bea
\epsilon^3&=&i(1-e_{1234})~,~~~\epsilon^4=-i k^2 u(1-e_{1234})+i h_\ui \Gamma^{+\ui} (1-e_{1234})~.
\la{su3ks}
\eea
The isotropy group of all these spinors is in $SU(3)$. To proceed, we find that a basis for the  1-form bi-linears is
\bea
\lambda^- &=& \bbe^-~,~~~
\lambda^+ =\bbe^+ - {1 \over 2} k^2 u^2 \bbe^- -u h~,~~~
\lambda^1 = k^{-1} \big(h+ k^2 u \bbe^-\big)~,
\cr
\lambda^6&=& k^{-1} \ell_\ui  \bbe^\ui~,~~~\ell_\ui=h_\uj {\bf I}^\uj{}_\ui~,
\la{su3con}
\eea
where the hermitian form of ${\bf I}$ is
\bea
\omega^{(8)}_{\bf I}=-(\bbe^1\wedge \bbe^6+\bbe^2\wedge \bbe^7+\bbe^3\wedge \bbe^8+\bbe^4\wedge \bbe^9)~.
\eea
All 1-form bi-linears  $\lambda^a$, $a=-,+, 1, 6,$ are  $\hat\nabla$-parallel.

As in the $G_2$ case, the associated vector fields $\xi_a$, $a=-,+,1$ to the 1-forms $\lambda^a$
span a $\mathfrak{sl}(2,\bR)$ Lie algebra. It remains to find
the commutator of $\xi_6$ with the other three vector fields. For this observe that
(\ref{ax1}) implies that $dh$ is orthogonal to all $\xi_a$ directions and
\bea
dh\in \mathfrak{su}(3)~.
\la{su3h}
\eea
In particular,
\bea
dh_{\ui\uj} \ell^\uj=0~.
\la{su3ort}
\eea
Moreover since all $\xi_a$ are $\hat\nabla$-parallel,
\bea
[\xi_6, \xi_a]=- i_{\xi_6} i_{\xi_a} H= i_{\xi_6} {\cal F}_a=0~, ~~~a=-,+,1~.
\la{6a}
\eea
The last equality follows from (\ref{su3ort}) and  (\ref{curvnh1}). Therefore,
the Lie algebra of the vector fields is
\bea
\mathfrak{sl}(2,\bR)\oplus \mathfrak{u}(1)~.
\eea
The heterotic horizons with Killing spinors that have isotropy group $SU(3)$
are principal bundles $M=P(SL(2, \bR)\times U(1),  B^6)$ with fibre group $SL(2,\bR)\times U(1)$
equipped with the connection $\lambda$ (\ref{su3con}). From the general results of \cite{het1}, the
spacetime metric and 3-form flux can be written as
\bea
ds^2=\eta_{ab}\lambda^a \lambda^b +d\tilde s^2_{(6)}~,~~~H=CS(\lambda)+\tilde H_{(6)}~,
\eea
where $d\tilde s^2_{(6)}$ and $\tilde H_{(6)}$ are orthogonal to the $\lambda^a$ directions.

To continue with the investigation of the geometry of spacetime, it remains to determine the geometry of $B^6$.
From the results of \cite{het1}, the spacetime admits a $\hat\nabla$-parallel  Hermitian  2-form $\omega$ and a (3,0)-form $\chi$
which are orthogonal to the $\lambda^a$ directions and are constructed as Killing spinor bi-linears. Moreover, the dilatino  KSE
implies that
\bea
\partial_a\Phi=0~,~~~\theta_\omega=2d\Phi~,~~~{\cal F}^6_{ij}\omega^{ij}=-2k~,~~~({\cal F}^6)^{2,0}=0~,~~~\tilde N(I)=0~,
\la{su3d}
\eea
where $I$ is the almost complex structure associated to $\omega$ and $ds^2_{(6)}$, the superscript in ${\cal F}^6$
denotes the (2,0) part of the curvature in a decomposition under $I$ in holomorphic and anti-holomorphic indices and $N$
is the Nijenhuis tensor of $I$.
Furthermore, (\ref{su3h}) implies that
\bea
{\cal F}^a_{ij} \omega^{ij}=0~,~~~({\cal F}^a)^{2,0}=0~,~~~a=-,+,1~.
\eea

The base space $B^6$ is a complex manifold. It admits a Hermitian form because
\bea
i_{\xi_a}\omega=0~,~~~{\cal L}_{\xi_a} \omega=0
\eea
and so $\omega$ descends to a Hermitian form on $B^6$. The last equation holds because $\hat\nabla\omega=0$, and the (2,0)-part
of all ${\cal F}$ curvatures vanishes. The integrability of the almost complex structures follows from the Nijenhuis condition
in (\ref{su3d}). This complex structure is compatible with geometric data $d\tilde s^2_{(6)}$ and $\tilde H_{(6)}$.
Furthermore observe that
\bea
i_{\xi_a} \chi=0~,~~~a=-,+,1, 6;~~~{\cal L}_{\xi_6} \chi= i k\chi~,~~~{\cal L}_{\xi_a} \chi=0~,~~~a=-,+,1~.
\eea
So although $\chi$ is orthogonal to the $\lambda^a$ directions, it is not invariant under the action of $\xi_6$ and so
it descends to $B^6$ as a line bundle valued  (3,0)-form. As a result $B^6$ does not have an $SU(3)$ structure but rather a $U(3)$ one.
In particular, the connection with skew-symmetric torsion, $\hat{\tilde \nabla}^{(6)}$, on $B^6$ has holonomy contained in $U(3)$. The torsion
can be expressed as
\bea
\tilde H_{(6)}=-i_I d\omega =e^{2 \Phi} \star_6 d [e^{-2\Phi} \omega]~,
\eea
see eg \cite{hull, howe, strominger, howegp2,  lust, waldram, salamon, ivanovy}.
The equation that remains to be solved is the closure of $H$. In particular, one finds that
\bea
dH=\eta_{ab} {\cal F}^a\wedge  {\cal F}^b+d\tilde H_{(6)}= k^{-2} dh\wedge dh+k^{-2} d\ell\wedge d\ell+ d\Big(e^{2 \Phi}\star_6 d [e^{-2\Phi} \omega]\Big)=0~.
\eea
As in the $G_2$ case, this is reminiscent of the equations that arise in the heterotic theory in the presence of anomalies.
However,
there are some differences. The connections that contribute to the anomaly term are abelian and one of them does not satisfy the
Hermitian-Einstein condition but rather it satisfies the Hermitian-Einstein condition with cosmological constant. In addition
$B^6$ does not have a $SU(3)$ structure but rather a $U(3)$ one. Moreover there is no relative minus sign between
the two terms which depend on the $h$ and $\ell$ in the ``anomaly'' term. Nevertheless there are sufficient
similarities between the above equations and the anomalous Bianchi identity which appears in the anomaly cancelation mechanism
to suggest that there may exist
solutions analogous to those found for the latter in \cite{liyau, fuyau, stefan2}.

\subsubsection{Geometry of ${\cal S}^8$}

 The first two  $\hat{\tilde \nabla}$-parallel spinors are as in (\ref{g2sks}). The additional two are
\bea
\eta_+^3=i(1-e_{1234})~,~~~\eta_-^4=i\Gamma^{+\ui}h_\ui (1-e_{1234})~.
\la{su3sks}
\eea
The isotropy group of both spinors in $Spin(8)$ is $SU(3)$. Therefore the holonomy of $\hat{\tilde \nabla}$ is contained in $SU(3)$.
The associated $\hat{\tilde \nabla}$-parallel bilinears are
\bea
h~,~~~\ell~,~~~\omega~,~~\chi~,
\eea
where $\ell$, $\omega$ and $\chi$ have the properties mentioned in the previous section. Both  $h$ and $\ell$ can be viewed as
the connections of a $T^2$ bundle over a base space $B^6$.  In fact ${\cal S}^8$ is a holomorphic $T^2$ fibration over $B^6$.
The complex structure on ${\cal S}^8$ is associated with the Hermitian form
\bea
\omega_{(8)}=k^{-2} h\wedge \ell+\omega~.
\eea
The  metric and torsion  of ${\cal S}^8$
can be written as
\bea
d\tilde s^2=k^{-2} (h\otimes h+\ell\otimes\ell)+d\tilde s_{(6)}^2~,~~~\tilde H= k^{-2}  (h\wedge dh+\ell\wedge d\ell)+\tilde H_{(6)}~,
\eea
where $d\tilde s_{(6)}^2$  and $\tilde H_{(6)}$ are given in previous section. Since $h$ and $\ell$ commute, one can adapt
coordinates such that
\bea
h=d\tau+p_i e^i~,~~~\ell=d\sigma+q_i e^i \ ,
\eea
where $p, q$, and all the other components of the fields, depend only on the coordinates of $B^6$.

\newsection{N=6 horizons}

\subsection{Geometry of Spacetime}

The first four Killing spinors are the same as those given for the $SU(3)$ case in (\ref{g2ks}) and (\ref{su3ks}). For the
additional two Killing spinors, it can be shown that they can be expressed as
\bea
\epsilon^5=i(e_{12}+e_{34})~,~~~\epsilon^6=-i k^2 u(e_{12}+e_{34})+i h_\ui \Gamma^{+\ui} (e_{12}+e_{34})~.
\la{6su2ks}
\eea

The isotropy group of all these spinors is in $SU(2)$. To proceed, we find that a basis in the space of
 1-form bi-linears is
\bea
\lambda^- &=& \bbe^-~,~~~
\lambda^+ =\bbe^+ - {1 \over 2} k^2 u^2 \bbe^- -u h~,~~~
\lambda^1 = k^{-1} \big(h+ k^2 u \bbe^-\big)~,
\cr
\lambda^{r'}&=& k^{-1} (\ell^{r'})_\ui\,  \bbe^\ui~,~~~(\ell^{r'})_\ui= -{({\bf I}^{r'})}^\uj{}_\ui h_\uj \ , \qquad r'=2,6,7
\la{su2con}
\eea
where ${\bf{I}}^{r'}$  is a triplet of almost complex structures
satisfying
the algebra of the imaginary unit quaternions
\bea
{\bf I}^{r'} {\bf I}^{s'}=-\delta^{r's'} 1_{8\times 8}+\epsilon^{r's'}{}_{t'} {\bf I}^{t'}~,~~\epsilon_{627}=1~.
\eea
 The associated Hermitian forms can be written
as
\bea
\omega^{(8)}_{{\bf{I}}^2} &=& \bbe^1\wedge \bbe^2+\bbe^3\wedge \bbe^4 + \bbe^{\bar{1}}\wedge \bbe^{\bar{2}}+ \bbe^{\bar{3}}\wedge \bbe^{\bar{4}}
\nn
\omega^{(8)}_{{\bf{I}}^6} &=& -i \big(\bbe^1\wedge \bbe^{\bar{1}}+ \bbe^2\wedge \bbe^{\bar{2}} + \bbe^3\wedge \bbe^{\bar{3}} +\bbe^4\wedge \bbe^{\bar{4}} \big)
\nn
\omega^{(8)}_{{\bf{I}}^7} &=&- i\big(\bbe^1\wedge \bbe^2+\bbe^3\wedge \bbe^4 - \bbe^{\bar{1}}\wedge \bbe^{\bar{2}}- \bbe^{\bar{3}}\wedge \bbe^{\bar{4}}\big) \ ,
\eea
in terms of the standard holomorphic frame basis.
These three hermitian forms also arise as spinor bilinears, which can be constructed explicitly
from the Killing spinors $\epsilon^1, \epsilon^3, \epsilon^5$, as described in Appendix A of \cite{het1}. Notice that
the isotropy group of $\epsilon^1, \epsilon^3, \epsilon^5$ is $Sp(2)\ltimes \bR^8$ and so this case here
is closely related to the backgrounds with 3 supersymmetries in \cite{het1}.

All 1-form bi-linears  $\lambda^a$, $a=+,-,1,2,6,7$ are  $\hat\nabla$-parallel, and so in particular their
associated vector fields $\xi_a$
are all Killing.
As in the $G_2$ case, the vector fields $\xi_a$, for  ($a=-,+,1$)
close under Lie brackets to a  $\mathfrak{sl}(2,\bR)$ Lie algebra. To find the rest of the commutators
first observe that (\ref{ax1}) implies that
\bea
dh\in \mathfrak{su}(2)~.
\la{su2h}
\eea
As a result, $dh$ is orthogonal to all $\xi_a$ directions. Then an argument similar to that which we have used in
equation (\ref{6a}) gives
\bea
[\xi_a, \xi_{r'}]=0~,~~a=-,+,1~.
\la{com}
\eea
It remains to find the commutators $[\xi_{r'}, \xi_{s'}]$. If  $[\xi_{r'}, \xi_{s'}]$ cannot be expressed
in terms of $\xi_a$, then one can show  there will be  additional linearly independent $\hat\nabla$-parallel vector
fields on the spacetime and therefore the holonomy of $\hat\nabla$ will be reduced to $\{1\}$. The only
such solutions are group manifolds and preserve 8 supersymmetries. Thus, we shall take
\bea
[\xi_{r'}, \xi_{s'}]=-c\, k\, \epsilon_{r's'}{}^{t'} \xi_{t'}~,
\la{6lie}
\eea
for some constant $c$. The above commutator cannot close in the $-,+$ and $1$ directions because of (\ref{com}), and  the structure
constants are skew.

Since the Killing spinors have isotropy group $SU(2)$, the holonomy of $\hat\nabla$ is contained in $SU(2)$.  It is then
a consequence of the results of \cite{het1, het2} that the spacetime is a principal bundle with fibre
the 6-dimensional Lorentzian group $G$
over a 4-manifold $B^4$ such that the spacetime metric and 3-form can be written as
\bea
ds^2&=& \eta_{ab} \lambda^a \lambda^b +e^{2\Phi}d\mathring s^2_{(4)}
\cr
H&=&CS(\lambda)+\tilde H_{(4)}~,
\eea
where we have re-scaled the 4-dimensional metric of the base space with the dilaton $\Phi$ for later convenience.
Moreover the spacetime
admits three $\hat\nabla$-parallel Hermitian forms $\omega^r$ which are orthogonal to all $\lambda^a$ 1-forms and
they are compatible with the metric $d\mathring s^2_{(4)}$.
The associated almost complex structures $I_r$ satisfy
\bea
I_r I_s=-\delta_{rs} 1_{4\times 4}+\epsilon_{rst} I_t~.
\eea
This is the full content
of the gravitino KSE.

To solve the dilatino KSE, first observe that another consequence of (\ref{su2h}) is that ${\cal F}^a$, $a=-,+, 1$ does not contribute.
This is because the Killing spinors are $SU(2)$ invariant and  ${\cal F}^a$, $a=-,+, 1$ is proportional to $dh$.
A direct computation of the remaining dilatino KSE using the $Sp(2)\ltimes \bR^8$ results of  \cite{het1}, or a
comparison with the backgrounds that preserve 6 supersymmetries in \cite{het3},  reveals
that
\bea
\partial_a\Phi=0~,~~~({\cal F}^{\rm sd})^{r'}=\nu \tilde \omega^{r'}~,~~~\nu={1\over2} (k+ck)~,~~\tilde H_{(4)}=-\mathring\star_{4} de^{2\Phi}~,
\la{6con}
\eea
where
\bea
({\cal F}^{\rm sd})^{r'}={1\over2}({\cal F}^{r'}+\star_{4}{\cal F}^{r'})~,~~~({\cal F}^{\rm ad})^{r'}={1\over2}({\cal F}^{r'}-\star_{4}{\cal F}^{r'})~,
\eea
are the self-dual and anti-self-dual components of ${\cal F}$, respectively, and $\omega^6\equiv \omega^1$,
$\omega^2\equiv \omega^2$ and $\omega^7\equiv \omega^3$.
 ${\cal F}^{\rm ad}$ is not restricted by the KSE.

There is an additional condition on the parameters of the solution. To see this first use $\hat\nabla \omega_{r'}=0$ and (\ref{6con})
to show that
\bea
{\cal L}_{r'}\omega_{s'}= 2\nu \epsilon_{r's't'} \omega_{t'}~.
\la{lieo}
\eea
Then either by comparing
$dH=0$ with the dilatino field equation or evaluating the identity $[{\cal L}_{\xi_{r'}}, {\cal L}_{\xi_{s'}}]={\cal L}_{[\xi_{r'},\xi_{s'}]}$
on $\omega^{t'}$  using (\ref{lieo}) and (\ref{6lie}), one finds that
\bea
(1+c) (2c+1)=0
\eea
and so either $c=-1$ or $c=-{1\over2}$. If $c=-1$, $\nu=0$ and the supersymmetry enhances to $N=8$. These solutions
will be investigated later.

Now, if $c=-{1 \over 2}$ these solutions are special cases of those given
in \cite{het3} that preserve 6 supersymmetries.
The only remaining equation that has to be solved is
\bea
 \mathring{\nabla}^2 e^{2\Phi}=-{1\over2} ({\cal F}^{\rm ad})_{ij}^{r'}({\cal F}^{\rm ad})^{ij}_{r'}-{k^{-2}\over2}
 dh_{ij} dh^{ij}+{3\over 8} k^2 e^{4\Phi}~,
\eea
where the inner products in the rhs have been taken with respect to the $d\mathring s^2_{(4)}$ metric.
The sign of the  rhs is indefinite.  As a result, there may be solutions which preserve strictly 6 supersymmetries.
As the $N=6$ solutions are included in the $N=2$ and $N=4$ heterotic horizons,
this sign is also significant for the existence of solutions in the $G_2$ and $SU(3)$ cases.

The geometry of $B^4$ can be investigated as in \cite{het3}. In particular, the self-dual component of the Weyl tensor of
$B^4$ vanishes but it may not be Einstein with cosmological constant. For the latter, the anti-self-dual part of the Weyl
tensor must vanish as well.
Because of (\ref{lieo}), the three hermitian forms $\omega_r$ do not descend on $B^4$ as hermitian
forms but they are rather twisted with an $SU(2)$ bundle. More details on the geometry of $B^4$ can be found in
\cite{het3}.

\subsubsection{Geometry of ${\cal S}^8$}

 The first four  $\hat{\tilde \nabla}$-parallel spinors are as in the $SU(3)$ case (\ref{g2sks}) and (\ref{su3sks}). The additional two Killing spinors
 can be written as
\bea
\eta_+^5=i(e_{12}+e_{34})~,~~~\eta_-^6=i\Gamma^{+\ui}h_\ui (e_{12}+e_{34})~.
\la{6su2sks}
\eea
The isotropy group of all 6 spinors in $Spin(8)$ is $SU(2)$. Therefore the holonomy of $\hat{\tilde \nabla}$ is contained in $SU(2)$.
The associated $\hat{\tilde \nabla}$-parallel bilinears are
\bea
h~,~~~\ell^{r'}~,~~~\omega_r~,
\eea
where $\ell^{r'}$ and  $\omega_r$ have the properties mentioned in the previous section. ${\cal S}^8$ is a $S^1\times S^3$ fibration
over $B^4$.  Both  $h$ and $\ell^{r'}$ can be viewed as
the connections of a $U(1)\times SU(2)$ bundle over a base space $B^4$.  The  metric and torsion  of ${\cal S}^8$
can be written as
\bea
d\tilde s^2=k^{-2} (h\otimes h+\sum_{r'}\ell^{r'}\otimes\ell^{r'})+e^{2\Phi}d{\mathring s}_{(4)}^2~,~~\tilde H= k^{-2}  (h\wedge dh)+CS(\lambda^{r'})
+\tilde H_{(4)}  ~ ,
\eea
where $d{\mathring s}_{(4)}^2$  and $\tilde H_{(4)}$ are given in the previous section.

\newsection{N=8 horizons}

The heterotic horizons that preserve 8 supersymmetries are special cases of the half supersymmetric backgrounds classified in
\cite{het4}. We shall demonstrate that within the class of $AdS_3$ horizons,  there are two heterotic horizon geometries with
8 supersymmetries which, up to discrete identifications,
are isometric to either $AdS_3\times S^3\times K_3$ or $AdS_3\times S^3\times T^4$.

The first six Killing spinors are as those of the solutions with 6 supersymmetries given in (\ref{g2ks}), (\ref{su3ks}) and
(\ref{6su2ks}). The remaining two can be chosen as
\bea
\epsilon^7&=&e_{12}-e_{34}~,~~~\epsilon^8=- k^2 u(e_{12}-e_{34})+h_\ui \Gamma^{+\ui} (e_{12}-e_{34})~.
\la{su2ks}
\eea
Since the first two  Killing spinors are those  that we have found in the $G_2$ case  (\ref{g2ks}),
the 1-form bilinears include those of (\ref{g2vbi}) for which the associated vector fields span the $\mathfrak{sl}(2,\bR)$ algebra. As a result
the fibre group of the spacetime has a $SL(2,\bR)$ subalgebra. It is then a consequence of the general classification results of
\cite{het1} that the spacetime is a principal bundle, $M=P(SL(2,\bR)\times SU(2), B^4)$,  with fibre group $AdS^3\times S^3$ and base space $B^4$ which
is conformal to a 4-dimensional hyper-K\"ahler manifold. Moreover the $AdS^3\times S^3$ fibre twists over the base space with the
connection $\lambda$ which is  an anti-self dual instanton. The dilaton $\Phi$ is function  of the base space $B^4$ and independent from all
the coordinates along the fibre directions.

The metric and 3-form flux can be written as
\bea
ds^2&=&\eta_{ab} \lambda^a \lambda^b+e^{2\Phi} d\mathring s_{(4)}^2~,
\cr
H&=&CS(\lambda^{\mathfrak{sl}})+CS(\lambda^{\mathfrak{su}})+\tilde H_{(4)}~,~~~~\tilde H_{(4)}=-\mathring\star_4\, de^{2\Phi}~,
\eea
where $d\mathring s_{(4)}^2$ is the hyper-K\"ahler metric on $B^4$, the Hodge operation in $\tilde H_{(4)}$ is taken with respect
to the hyper-K\"ahler metric and $a,b=0,5,1,6,2,7$. Moreover, we have split the Chern-Simons
form of $\lambda$ into the $\mathfrak{sl}(2, \bR)$ and $\mathfrak{su}(2)$
parts.

The only equation that remains to be solved is the closure of $H$, $dH=0$. This in turn implies, using the anti-self duality of the curvature
${\cal F}$ of $\lambda$, that
\bea
\mathring\nabla^2 e^{2\Phi}=-{1\over2} \eta_{ab} {\cal F}^a_{ij} {\cal F}^{bij}~,
\la{su2eqnm}
\eea
where $i,j=3,8,4,9$ are now $B^4$ indices and the indices in the right-hand-side have  been  raised with respect to the hyper-K\"ahler metric. The right-hand-side of
(\ref{su2eqnm}) has a definite sign. The possibility of an indefinite sign only arises from the $\mathfrak{sl}(2,\bR)$
component of the curvature ${\cal F}$. But a straightforward calculation reveals, using (\ref{sl2rc}), that
\bea
\eta_{ab} {\cal F}^a_{ij} {\cal F}^{bij}\vert_{\mathfrak{sl}(2,\bR)}= k^{-2}(dh)_{ij} (dh)^{ij}\geq 0~.
\eea
Thus the right-hand-side of
(\ref{su2eqnm}) is strictly negative and so a partial integration argument over the compact base space $B^4$ implies
that it should vanish. Thus we conclude that for the horizon geometries
\bea
{\cal F}^a=0~,
\eea
and the dilaton $\Phi$ is constant. Since the curvature of the principal bundle vanishes, the spacetime is a product\footnote{If $B^4$ is not simply connected, there is the possibility that the fibre twists
with a flat but not trivial connection $\lambda$.  We shall not investigate this further here.} $AdS_3\times S^3\times B^4$, where $B^4$ is a compact
4-dimensional hyper-K\"ahler manifold
and so it is either $T^4$ or $K_3$.

\newsection{Non-vanishing anomaly and near brane horizons}

So far in our analysis, we have not included the heterotic anomaly contribution. In particular,  we have assumed that the gravitational anomaly
cancels the gauge sector anomaly, and so the 3-form flux is closed. If the anomaly contributes, our analysis is
modified. There are two ways of thinking about this. One is in the context of perturbation theory where the KSEs, the field
equations and  $dH$ receive higher order curvature corrections expressed as a Taylor series expansion in $\alpha'$.
Therefore one expects that the spacetime metric and 2-form gauge potential for generic backgrounds get corrected to all orders in perturbation theory.
 Alternatively,  one can take the anomaly contribution which appears at order $\alpha'$ as exact. Such an assumption
  has led  to  applications
in differential geometry \cite{liyau}. These two different ways of thinking give distinct results which we shall explain. We shall mostly
focus on perturbation theory and then we shall comment how our analysis is altered in the exact case.

\subsection{Perturbation theory}

We shall not explore the theory to all orders in $\alpha'$. Instead we shall follow \cite{ptsimp} and consider the correction up to and including
two loops in sigma model perturbation theory \cite{hulltown}. Now taking $dH\not=0$,
it is known that the KSEs of supergravity theory do not change  up to this correction \cite{bergshoeff}.
However field equations do.

We are working in  perturbation theory and so a background gets corrected
order by order in $\alpha'$. Since the anomaly contributes in the first order in $\alpha'$, we must  begin
 from a configuration for which the 3-form field strength is closed. As a result,
the analysis we have performed is valid at the zeroth order in $\alpha'$. Moreover since the Killing spinor
equations remain the same up to the order we consider, the only alteration in the analysis\footnote{There is the possibility
that the anomaly correction is not compatible with the Einstein equation when the two loop correction is included. This
has been examined in detail in \cite{ptsimp, stefan2}. }  is to set
\bea
dH=-{\alpha'\over4} [{\rm tr} \check R\wedge \check R-{\rm tr} (F\wedge F)]+{\cal O}(\alpha'^2)~,
\la{anbianchi}
\eea
instead of taking $dH=0$ as in previous sections, where the contribution in the rhs is due to the anomaly, $\check R$ is the curvature
of the connection $\check \nabla=\nabla-{1\over2} H$ and $F$ is the curvature of the gauge sector. Observe that since the rhs is
first order in $\alpha'$, the contribution
in $\check R$ comes from the zeroth order metric and 2-form gauge potential and so
\bea
\hat R_{AB,CD}=\check R_{CD,AB}+{\cal O}(\alpha')~,
\la{hcb}
\eea
as a consequence of a Bianchi identity and $dH=0+{\cal O}(\alpha')$.
Therefore, $\check \nabla$ is an $\mathfrak{Lie}({\rm hol}(\hat\nabla))$-instanton connection with gauge group contained in $SO(9,1)$, ie
\bea
\check R\in \mathfrak{Lie}({\rm hol}(\hat\nabla))~,
\eea
where $\mathfrak{Lie}({\rm hol}(\hat\nabla))= G_2, ~ SU(3)$ or $SU(2)$.
The gaugino KSE also implies that $F$ is  a  $\mathfrak{Lie}({\rm hol}(\hat\nabla))$-instanton and the gauge group must be a subgroup
of $E_8\times E_8$ or $Spin(32)/\bZ_2$. In (\ref{anbianchi}), $\check R$  can be replaced with any
other curvature of the spacetime because of the scheme dependence of perturbation theory. However
 to preserve
 extended worldvolume supersymmetry in perturbation theory, $\check R$ is a natural choice \cite{howepapad}.

First suppose that we are considering the heterotic horizons $\bR^{1,1}\times {\cal S}^8$ tabulated in table 1.
Since the zeroth order contribution for $H$ vanishes, $\check R=R$. The solutions, although they begin with vanishing
torsion, can develop non-vanishing torsion which is proportional to $\alpha'$. This is the case for all backgrounds
preserving up to and including 8 supersymmetries as demonstrated in \cite{ptsimp}. Moreover, it is expected that there will be corrections
to all order in $\alpha'$. The precise corrections that they receive depend on a case by case analysis.

Next let us consider heterotic horizons associated with $AdS_3$. In this case the anomalous Bianchi identity
(\ref{anbianchi}) can be rewritten as
\bea
d\tilde H_{(n)}=-\eta_{ab} {\cal F}^a\wedge {\cal F}^b-{\alpha'\over4} [{\rm tr} \check R\wedge \check R-{\rm tr} (F\wedge F)]+{\cal O}(\alpha'^2)~,
\la{anbianchib}
\eea
where $\tilde H_{(n)}$ is the 3-form field strength of the base space and so $n=7,6$ or $4$. However since the zeroth order in $\alpha'$ term in
$H$ does not vanish, $\check R$ is different from $R$. It is expected that there will be $\alpha'$ corrections to the
fields for all backgrounds that preserve less than 8 supersymmetries.

To investigate the case with 8 supersymmetries first observe that the first term in the rhs of (\ref{anbianchib}) vanishes  because ${\cal F}=0$.
Then (\ref{anbianchib}), following
(\ref{su2eqnm}), can be rewritten as
\bea
\mathring\nabla^2 e^{2\Phi}=-{\alpha'\over 8} {\rm tr}\check R_{ij} \check R^{ij}+
{\alpha'\over 8}{\rm tr} F_{ij} F^{ij}+{\cal O}(\alpha'^2)~.
\la{su2eqnm2}
\eea
The gaugino KSE
requires that $F$ is an anti-self-dual instanton on $B^4$. It is significant that the inclusion of the anomaly makes
the sign of the rhs indefinite. Therefore solutions with a non-trivial dilaton cannot be ruled out. In
the $AdS_3\times S^3\times K_3$ background, $\check R=R\not=0$ and  so the rhs of (\ref{su2eqnm2}) may not vanish. The existence
of $\alpha'$ corrections depends on the choice of $F$.  However for the $AdS_3\times S^3\times T^4$ background $\check R=0$, then using
the compactness of $T^4$ one concludes that $F=0$ and $\Phi$ is constant at order $\alpha'$. Therefore $AdS_3\times S^3\times T^4$ is an exact
solution up to  and including 2 loops in the sigma model perturbation theory. Since it is a group manifold solution, it must be exact to all
orders in perturbation theory.

\subsection{Exact modification}

Next suppose that the $\alpha'$ correction to $dH$ is exact.
The Bianchi identity of $H$ is different from that of (\ref{anbianchi}) and reads
\bea
dH=-{\alpha'\over4} [{\rm tr} \dot R\wedge \dot R-{\rm tr} (F\wedge F)]~,
\la{anbianchic}
\eea
where now $\dot R$ is a curvature of the spacetime which is a $\mathfrak{Lie}({\rm hol}(\hat\nabla))$-instanton with gauge group
$SO(9,1)$ and similarly for $F$. Moreover $\alpha'$ is any  positive number. Since $dH\not=0$, $\check R$ is not a
$\mathfrak{Lie}({\rm hol}(\hat\nabla))$-instanton and so it cannot be identified with $\dot R$.

Consistency requires that
the field equations are also modified. In particular both the Einsten and dilaton field equations alter.
In particular the dilaton field equation now reads
\be
\tilde \nabla^2 \Phi -2 \tilde \nabla^A \Phi \tilde \nabla_A \Phi +{1 \over 12}  H_{ABC}  H^{ABC}
+{\alpha'\over 8} {\rm tr}\dot R_{AB} \dot R^{AB}
-{\alpha'\over 8}{\rm tr} F_{AB} F^{AB}  =0~.
\ee
Note that the sign of the curvature terms associated with the anomaly is indefinite, and so the arguments
presented in sections 3 and 4 for the heterotic horizons do not generalize. Therefore a new investigation is required.
It is likely  that many more solutions exist in this case as the gauge connection can be used to lessen
the restrictions on the differential system imposed by the compactness of ${\cal S}^8$.

\subsection{Brane horizons and dilaton singularities}

Some of the heterotic horizon geometries we have found  can be identified as the near horizon geometries of brane configurations. To distinguish between
the two, we shall refer to the latter as ``near brane geometries''.
In particular the near horizon geometry   $AdS_3\times S^3\times T^4$  arises as the near brane geometry
of a fundamental string \cite{gibbonsruiz} on a 5-brane \cite{callan}. In this identification, $AdS_3$ is spanned by the worldvolume
directions of the string and the overall radial transverse direction, $S^3$ is the 3-sphere of the overall transverse sphere and $T^4$ are the relative
transverse directions of the string on the worldvolume of the 5-brane. This is also the case with the $AdS_3\times S^3\times K_3$ solution.
Again one should consider a fundamental string on a 5-brane but replace the $T^4$ relative transverse directions of the
string on the 5-brane with $K_3$. So both heterotic horizons that preserve 8 supersymmetries can arise as near brane geometries.

Some brane configurations in heterotic supergravity exhibit  near brane geometries
which are not included in the analysis we have done for the heterotic horizons.  For example
the near brane geometry of the 5-brane is $\bR^{1,1}\times T^4\times S^3\times S^1$ and has a
linear dilaton depending on the angular coordinate of $S^1$ \cite{callan}.  Another similar example is  $\bR^{1,1}\times T^2\times S^3\times S^3$
with linear dilaton depending on  the angular coordinates of $T^2$, and arises as the near brane geometry of
the 5-brane configurations of \cite{gptesc}.

There are two main differences between the near brane geometries mentioned above and the near horizon geometries that we
have investigated. First, both $\bR^{1,1}\times T^4\times S^3\times S^1$ and $\bR^{1,1}\times T^2\times S^3\times S^3$ near
brane geometries are not horizons but rather other asymptotic regions. They are located at infinite affine distance away from
any interior point of the brane spacetime. Another difference is that the dilaton is not well defined
on the near brane spacetime because it is not periodic in an angular coordinate(s). As a result, it cannot be thought of
as a well-defined function of the compact section of the near brane geometry. It is clear from both these points that the
KSEs do not guarantee either analyticity in the radial direction $r$ or regularity of the fields on the horizon section. However,
enforcing one or the other will rule both these near brane geometries out. As a result, it may  be that enforcing one
of them, together with the Killing spinor  and field equations, will imply the other.

\newsection{Concluding Remarks}

We have found that there are two classes of heterotic horizons. One class is $\bR^{1,1}\times {\cal S}^8$, where
${\cal S}^8$ is a product of special holonomy manifolds which admit parallel spinors, the dilaton is constant and
the 3-form flux vanishes. A more detailed description is given in table 1. The other class of solutions contains
$AdS_3$ as subspace and preserves 2, 4, 6 and 8 supersymmetries. In particular, the spacetime is a fibration
 with fibre that contains $AdS_3$, and $AdS_3$ twists over the base space with a suitable $U(1)$ connection.
The solutions with 8 supersymmetries are  isometric up to discrete identifications with
$AdS_3\times S^3\times K_3$, $AdS_3\times S^3\times T^4$ or  $\bR^{1,1}\times T^4\times K_3$, the radii of $AdS_3$ and $S^3$
are equal and the dilaton is constant. Clearly the first two heterotic horizons are of the $AdS_3$ class while the third
is of the $\bR^{1,1}\times {\cal S}^8$ class. Moreover, we have shown that $AdS_3\times S^3\times T^4$ does not receive
$\alpha'$ corrections.

Throughout most of our analysis, we have taken $dH=0$, so our results are automatically extended to the common sector of type II supergravities.
Clearly  all the heterotic horizons of table 1 can be interpreted as solutions of the
common sector preserving  twice as many  supersymmetries. The $AdS_3$ class of heterotic horizons, without the anomaly correction,
 can also be embedded
in the common sector of type II supergravities but it is not a priori apparent   that there will be a doubling of supersymmetry.
An exception to this is the heterotic   horizons which preserve 8 supersymmetries. It can be easily seen that as common sector solutions they
preserve 16 supersymmetries. Moreover these are the only common sector horizons which preserve 16 supersymmetries. This is because
if the left or right sector KSEs preserve more than 8 supersymmetries, then the solutions are plane waves \cite{jose2, het2}
and in particular  they
do not contain $AdS_3$ as a subspace. So for the common sector horizons to preserve 16 supersymmetries, the left and right sector
KSEs must preserve precisely 8 supersymmetries each. This proves that the common sector horizon geometries
that preserve 16 supersymmetries are locally isometric to $AdS_3\times S^3\times K_3$, $AdS_3\times S^3\times T^4$ or  $\bR^{1,1}\times T^4\times K_3$.

Having determined the geometry of heterotic horizons, it is natural to wonder whether
there are extreme black holes which exhibit such near horizon geometries. It is not a priori clear that this is the case
for all heterotic horizons that we have found.
However the solutions which contain an $AdS_3$ subspace can be trivially identified with a Kaluza-Klein black hole. This is
because they can be seen as an embedding of the 3-dimensional  black hole \cite{BTZ} in the heterotic supergravity. It is also known that
the near horizon geometries of black rings have an  $AdS_3$ subspace. So alternatively,
it may be possible to view this type of heterotic horizon as the near horizon geometry of a
 Kaluza-Klein black ring.

\vskip 0.5cm
{\bf Acknowledgments:}~We would like to thank Gary Gibbons, Ulf Gran,  Niels Obers, Harvey Reall, David Robinson and  Andrew Swann
for correspondence and helpful
discussions. JG is supported by the EPSRC grant EP/F069774/1. GP is
 partially supported
by the EPSRC grant EP/F069774/1 and the STFC rolling grant ST/G000/395/1.

\setcounter{section}{0}
\setcounter{subsection}{0}

\appendix{ Supersymmetric horizons}

The main purpose of the appendix is to show that if we take Gaussian null coordinates with respect
to the null Killing vector field constructed as a Killing spinor bi-linear, then, without loss of generality,
we can make the identification (\ref{ebbe}) leading to (\ref{shh}).

Before proceeding with the analysis of the KSEs, note that
the non-vanishing components of the spin connection are given by
\bea
&&\Omega_{+,-i} = -{1 \over 2} h_i~, \qquad \Omega_{+,ij} = -{1 \over 2} r (dh)_{ij}~,\qquad \Omega_{-,+i} = -{1 \over 2} h_i~,
\cr
&&\Omega_{i,+-} ={1 \over 2} h_i~, \qquad \Omega_{i,+j} = -{1 \over 2} r (dh)_{ij}~,
\qquad \Omega_{i,jk} = \tilde\Omega_{i,jk}~,
\la{spincon}
\eea
where $\tilde\Omega$ is the spin connection of the horizon section ${\cal S}^8$.

\subsection{ Gravitino}

The gravitino equation $\hat\nabla\e=0$ can be written explicitly as
\be
 \partial_A\e + {1 \over 4} \Omega_{A, B_1 B_2} \Gamma^{B_1 B_2}\e - {1 \over 8} H_{A B_1 B_2} \Gamma^{B_1 B_2} \epsilon=0~,
\ee
where the spin connection is given in (\ref{spincon}) and $H$ in (\ref{bhdata}).
To proceed, we write
\be
\epsilon = \epsilon_+ + \epsilon_-~,~~~\Gamma_\pm \epsilon_\pm =0~.
\ee
After some straightforward computation, it is possible to solve the $+$ and $-$ components of
the KSE to find
\bea
\epsilon_+ &=& \eta_+ + {1 \over 4} u \big( (S-1)h-dS+N \big)_i \Gamma^i \Gamma_+ \eta_-~,
\cr
\epsilon_- &=& \eta_- -{1 \over 4} r \big((S+1)h-dS+N \big)_i \Gamma^i \Gamma_- \eta_+
\cr &&~~~~~~~~~~~
-{1 \over 8} u r\big( (S-1)dh - h \wedge N + dN \big)_{ij} \Gamma^{ij} \eta_-~,
\label{ksp1}
\eea
where $\eta_\pm = \eta_\pm (y)$ do {\it not} depend on $r$ and $u$. These expressions are {\it sufficient} to prove
(\ref{ebbe}) but we shall state the rest of the equations arising from the KSEs for completeness.

In addition, the $+$ and $-$
components of the KSE   give

\bea
\label{al1}
\big( -(Sh-dS+N)^2+h^2 + d((S-1)h+N)_{ij} \Gamma^{ij} \big) \eta_+ =0~,
\eea
\bea
\label{al2}
\big((S-1)dh-h \wedge N + dN \big)_{ij} \Gamma^{ij} \big((S+1)h-dS+N \big)_\ell \Gamma^\ell \eta_+=0~,
\eea
\bea
\label{al3}
\big( (Sh-dS+N)^2-h^2 + d((S-1)h+N)_{ij} \Gamma^{ij} \big) \eta_- =0~,
\eea
\bea
\label{al4}
\bigg( \big((S-1)h-dS +N \big) \wedge \big((S-1)dh - h \wedge N + dN \big) \bigg)_{\ell_1 \ell_2 \ell_3} \Gamma^{\ell_1
\ell_2 \ell_3} \eta_-=0~,
\eea
\bea
\label{al5}
&&\bigg( - ((S-1)dh - h \wedge N + dN)_{ij} ((S-1)dh - h \wedge N + dN)^{ij}
\cr
&&+{1 \over 2} ((S-1)dh - h \wedge N + dN)_{\ell_1 \ell_2} ((S-1)dh - h \wedge N + dN)_{\ell_3 \ell_4}
\Gamma^{\ell_1 \ell_2 \ell_3 \ell_4} \bigg) \eta_-=0~. \nonumber \\
\eea

The remaining components of the KSE then imply that
$\eta_\pm$ satisfy

\be
\label{dif1}
\hn_i \eta_\pm + \big(-{1 \over 8} (dW)_{ijk} \Gamma^{jk}
\pm {1 \over 4}(-(S+1)h+dS-N)_i \big) \eta_\pm =0~,
\ee

together with the following algebraic constraints
\bea
\label{al6}
&&\bigg(- \hn_i((S+1)h-dS+N)_j -{1 \over 2}((S+1)h-dS+N)_j((S-1)h-dS+N)_i
\nn
&&+ \big((S+1)dh - h \wedge N + dN\big)_{ij} +{1 \over 2}((S+1)h-dS+N)^\ell (dW)_{ij \ell} \bigg) \Gamma^j \eta_+=0
\eea

\bea
\label{al7}
&& \bigg( \big(- \hn_i \big((S-1)dh-h \wedge N + dN\big)_{\ell_1 \ell_2}
- \big((S-1)dh-h \wedge N+dN\big)_{[\ell_1}{}^m(dW)_{|im| \ell_2]}
\nn
&& + h_i \big((S-1)dh-h \wedge N+dN\big)_{\ell_1 \ell_2} \big) \Gamma^{\ell_1 \ell_2}
\nn
&&- \big((S+1)dh -h \wedge N + dN\big)_{ij} \big((S-1)h-dS+N\big)_\ell \Gamma^j \Gamma^\ell
\bigg) \eta_-=0~,
\eea

\bea
\label{al8}
&&\bigg( \hn_i \big((S-1)h-dS+N\big)_j +{1 \over 2}\big((S-1)h-dS+N\big)_j
\big(-(S+1)h+dS-N\big)_i
\nn
&&-{1 \over 2} \big((S-1)h-dS+N\big)^\ell (dW)_{ij\ell} \bigg) \Gamma^j \eta_-=0~.
\eea

\subsection{ Dilatino }

To proceed further, we consider the dilatino KSE

\bea
\label{dileq}
\bigg( \partial_A \Phi \Gamma^A -{1 \over 12} H_{B_1 B_2 B_3}\Gamma^{B_1 B_2 B_3} \bigg) \epsilon =0~.
\eea

On substituting ({\ref{ksp1}}) into ({\ref{dileq}}),
one finds the following algebraic conditions

\bea
\label{al9}
\bigg( (2 d \Phi +dS-N-Sh)_i \Gamma^i -{1 \over 6} (dW)_{i_1 i_2 i_3} \Gamma^{i_1 i_2 i_3}\bigg) \eta_+=0~,
\eea
\bea
\label{al10}
\big(h \wedge N - dN -S dh\big)_{\ell_1 \ell_2} \Gamma^{\ell_1 \ell_2} \eta_+=0~,
\eea
\bea
\label{al11}
\big(h \wedge N - dN -S dh\big)_{\ell_1 \ell_2} \Gamma^{\ell_1 \ell_2} \big((S+1)h-dS+N \big)_i \Gamma^i \eta_+=0~,
\eea
\bea
\label{al12}
\bigg( (2 d \Phi +dS-N-Sh)_i \Gamma^i -{1 \over 6} (dW)_{i_1 i_2 i_3} \Gamma^{i_1 i_2 i_3}\bigg) \big((S-1)h-dS+N\big)_\ell
\Gamma^\ell \eta_-=0~,
\eea
\bea
\label{al13}
\big(h \wedge N - dN -S dh\big)_{\ell_1 \ell_2} \Gamma^{\ell_1 \ell_2} \big((S-1)dh-h \wedge N+dN\big)_{i_1 i_2}
\Gamma^{i_1 i_2} \eta_-=0~,
\eea
\bea
\label{al14}
\bigg( (2 d \Phi -dS+N+Sh)_i \Gamma^i -{1 \over 6} (dW)_{i_1 i_2 i_3} \Gamma^{i_1 i_2 i_3}\bigg) \eta_-=0~,
\eea
\bea
\label{al15}
\bigg( (2 d \Phi -dS+N+Sh)_i \Gamma^i -{1 \over 6} (dW)_{i_1 i_2 i_3} \Gamma^{i_1 i_2 i_3}\bigg)
\big((S+1)h-dS+N\big)_j \Gamma^j \eta_+=0~,
\eea
\bea
\label{al16}
&&\bigg( (2 d \Phi -dS+N+Sh)_i \Gamma^i -{1 \over 6} (dW)_{i_1 i_2 i_3} \Gamma^{i_1 i_2 i_3}\bigg)
\nn
&& \times \big((S-1)dh-h \wedge N + dN\big)_{q_1 q_2} \Gamma^{q_1 q_2} \eta_-=0~.
\eea

\subsection{N=1 Supersymmetry}

Now let us focus on the solutions that preserve one supersymmetry for which ${\partial \over \partial u}$ is identified with the
Killing spinor bi-linear. The components of the associated 1-form bilinear $X\equiv \bbe^-$ in the  basis ({\ref{nhbasis}}) are
\be
X_+ = 0 , \qquad X_-=1, \qquad X_i=0 \ .
\ee
It is clear that the data, including the KSEs, are invariant under local y-dependent $Spin(8)$ gauge transformations.
As $\eta_\pm$ depend only on $y$, there is a gauge transformation such that
\bea
\eta_+=\alpha(y) (1+e_{1234})~,~~~\eta_-=\beta(y) (e_{15}+e_{2345})~.
\la{guagsp}
\eea
This follows from the fact that $Spin(8)$ has a single type of orbit in the Majorana-Weyl 8-dimensional
representation with isotropy group $Spin(7)$, and $Spin(7)$ has also a single type of orbit in the anti-Majorana-Weyl 8-dimensional
representation with isotropy group $G_2$.

Next, consider the spinor bilinear
\be
Y=Y_A\, \bbe^A\equiv  \langle B \epsilon^*, \Gamma_A \epsilon \rangle\, \bbe^A~,
\ee
using (\ref{ksp1}) and (\ref{guagsp}).
By comparing the components of $Y$ with those of $X$ in the basis ({\ref{nhbasis}}),
we require that
\be
Y_+|_{r=0} =0
\ee
which imposes the condition $\beta=0$, i.e.
\be
\eta_-=0~.
\ee

Next, we require that the $O(r)$ term in the spinor bilinear $Y$ should vanish.
This imposes the condition
\be
\langle B(1+e_{1234}) , \Gamma_M ((S+1)h - dS + N)_i \Gamma^i  \Gamma_- (1+e_{1234} \rangle =0~,
\ee
for all $M$. This implies that
\be
(S+1)h - dS + N =0~.
\ee
Hence
\be
\epsilon = \eta_+ = \alpha(1+e_{1234}) \ .
\ee
Next, observe that
\be
Y_- = -2 \sqrt{2} \alpha^2 \ .
\ee
On comparing with $X_-=1$, we require that $\alpha$ be constant; without loss of generality take
$\alpha=1$, so
\be
\epsilon= \eta_+ = 1+e_{1234}~.
\ee

Having obtained this simplification, it is straightforward to summarize the conditions for the $N=1$ heterotic horizon
geometries as

\be
\label{cc1}
\hn_i \eta_+ -{1 \over 8} (dW)_{ijk} \Gamma^{jk} \eta_+ =0~,
\ee
\be
\label{cc2}
(S+1)h -dS + N =0~,
\ee
\be
\label{cc3}
(dh)_{ij} \Gamma^{ij} \eta_+ =0~,
\ee
\be
\label{cc4}
\bigg((2 d \Phi + h)_i \Gamma^i -{1 \over 6} \Gamma^{ijk} (dW)_{ijk} \bigg) \eta_+ =0~.
\ee

Note that ({\ref{cc2}}) can be used to eliminate $S$ and $N$ from the 3-form field strength $H$ to obtain

\be
H =\bbe^+ \wedge \bbe^- \wedge h + r \bbe^+ \wedge dh + dW~.
\ee
 Thus we have justified (\ref{ebbe}) and derived (\ref{shh}).


\end{document}